\documentstyle[12pt]{article}

\newcommand{\Od}{{\cal O}}

\newcommand{\lsim}   {\mathrel{\mathop{\kern 0pt \rlap
  {\raise.2ex\hbox{$<$}}}
  \lower.9ex\hbox{\kern-.190em $\sim$}}}
\newcommand{\gsim}   {\mathrel{\mathop{\kern 0pt \rlap
  {\raise.2ex\hbox{$>$}}}
  \lower.9ex\hbox{\kern-.190em $\sim$}}}

\begin{document}
\input epsf \renewcommand{\topfraction}{0.8}
\pagestyle{empty} \vspace*{5mm}
\begin{center}
\Large{\bf Cosmological and astrophysical limits on brane
fluctuations}
\\ \vspace*{2cm}
\large{ J. A. R. Cembranos, A. Dobado and A. L. Maroto}
\\ \vspace{0.2cm} \normalsize
Departamento de  F\'{\i}sica Te\'orica,\\
 Universidad Complutense de
  Madrid, 28040 Madrid, Spain\\
\vspace*{0.6cm} {\bf ABSTRACT} \\
  
\end{center} We consider a general brane-world model parametrized by the 
brane
tension scale $f$ and the branon mass $M$. For low tension compared to
the fundamental gravitational scale, we calculate the relic branon abundance
and its contribution to the cosmological dark matter. We compare this
result with the current observational limits on the total and hot dark matter
energy densities and derive the corresponding bounds on $f$ and $M$. Using the
nucleosynthesis bounds on the number of relativistic species, we also set a 
limit on the number of light branons in terms of the brane tension. Finally,
we estimate the bounds coming from the energy loss rate in supernovae 
explosions due to massive branon emission.
\vspace*{5mm}

\noindent
%\vspace*{0.5cm}
\begin{flushleft} PACS: 95.35.+d, 11.25-w, 11.10Kk \\
\end{flushleft}
%\vspace*{0.5cm}
%\pagestyle{empty}
%\clearpage\mbox{}\clearpage
\newpage
\setcounter{page}{1} \pagestyle{plain}
%\baselineskip 0.83 true cm
\textheight 20 true cm
\section{Introduction}

The increasing observational precision is making cosmology a 
useful tool in probing certain properties of particle physics
theories beyond the Standard Model (SM). The limits on the 
neutrino masses and the number of 
neutrino families are probably the most clear examples 
\cite{WMAP,Steigman}.   
 In some cases, the cosmological bounds are complementary to
those obtained from colliders experiments and therefore, the 
combination of both allows us to restrict the parameter space of a 
theory in a more efficient way.

There are two main ways in which cosmology can help. On one hand we
have  the relative abundances of the light elements, which is one 
of the most 
precise predictions of the standard cosmological
model. Indeed, the calculations are very sensitive
to certain cosmological parameters and thus for instance, the production 
of  $^4$He increases with the rate of expansion of the 
universe $H$. A succesful nucleosynthesis requires that $H$ should not
deviate from its standard value  more than around a 10$\%$ during 
that epoch.
Since $H$ depends on the  effective number of 
relativistic degrees of freedom $g_{eff}(T)$ at a given temperature
$T$, the above constraint translates into a bound on  
$g_{eff}(T_{nuc})$, where $T_{nuc}\sim 1 \mbox{MeV}$. Apart from the
number of particle species, cosmology also sets a limit on their 
energy density. For particles which are non relativistic 
at present, there exists an upper bound given by the measured 
dark matter density $\Omega_M=0.23\pm 0.08$ at the 95$\%$ C.L \cite{WMAP}.  

On the other hand, stars also provide useful information
on theories containing light and weakly interacting particles.
The extreme opacity of ordinary matter to photons makes it
very long the time it takes for a photon produced in the center
of the star to reach its surface. Indeed, this fact explains
the longevity of stars.  However, particles like
neutrinos or axions, still can be produced abundantly in nuclear reactions
in the core of the star, but since they are weakly interacting, the rate at
which they can carry away energy can be much larger. This was particularly
evident in the 1987A supernova explosion, in which most of the energy 
was released in the form of neutrinos. Again, this fact can be
used to set limits on the mass and couplings of the new particles.

In this paper we will study the constraints that cosmology and
astrophysics impose on the so called brane-world scenario (BWS), which
is becoming one of the most popular extensions of the SM. In these
models, the Standard Model particles are bound to live on a 
three-dimensional brane embedded in a higher dimensional ($D=4+N$) 
space-time,
whereas gravity is able to propagate in the whole bulk space. 
The fundamental
scale of gravity  in $D$ dimensions $M_D$  can be lower than the 
Planck scale $M_P$. In
the original proposal in \cite{ADD}, the main aim was to address the
hierarchy problem, and for that reason the value of $M_D$ was taken
around the electroweak scale. However more recently, brane 
cosmology models have been proposed in which $M_D$ has to be much larger
than the TeV \cite{Langlois,six}. In this work we will consider a general BWS with arbitrary
fundamental scale $M_D$.

The existence of extra dimensions is responsible for the appearence of
new fields on the brane. On one hand, we have the tower of Kaluza-Klein
modes of fields propagating in the bulk space, i.e. the gravitons. 
On the other, since the brane has a finite tension $f^4$, its 
fluctuations will be parametrized by some $\pi^\alpha$ fields 
called branons. These fields, in the case in which traslational invariance
in the bulk space is an exact symmetry, can be understood as the massless
Goldstone bosons arising from the spontaneous breaking of that symmetry
induced by the presence of the brane \cite{Sundrum,DoMa}.
However, in the most general case, translational invariance will
be explicitly broken and therefore we expect branons to be massive
fields. 

It has been shown \cite{GB} 
that when 
branons are properly taken into account, the coupling of the SM
particles to any bulk field is exponentially suppressed by a
factor $\exp(-M^2_{KK}M_D^2/(8 \pi^2f^4))$, where $M_{KK}$ is the
mass of the corresponding KK mode. As a consequence, if the tension
scale $f$ is much smaller than the fundamental scale $M_D$, i.e.
$f\ll M_D$, the KK modes decouple from the SM particles. Therefore, 
for flexible enough branes, the only relevant degrees of freedom at
low energies in the BWS are the SM particles and branons.

The phenomenological implications of KK gravitons for colliders 
physics, cosmology and astrophysics have been studied in a series of
papers (see \cite{KK} and references therein), 
and the corresponding limits on $M_D$ and/or the number of
extra dimensions $N$ have been obtained.  
In the case of branons, the
potential signatures in  colliders have been studied in the massless
case in \cite{strumia} and in the massive one in \cite{ACDM}. Limits
from supernovae and modifications of Newton's law at small distances 
in the massless case were obtained in \cite{Kugo}. Moreover 
in \cite{CDM} the interesting 
possibility that massive branons could account for the observed dark matter
of the universe was studied in detail. The main aim of the paper is 
to analyse the cosmological and astrophysical limits on the BWS through
the effects due to massive branons. 
To that end we will be
assuming  that  the evolution of
the universe is standard up to a temperature around $f$.
Indeed, this is the case of realistic brane cosmology models
in five dimensions \cite{Langlois}. Also more recently, six-dimensional 
models have been proposed in which the Friedmann equation has the 
standard form for arbitrary temperature \cite{six}.

The paper is organized as follows: in 
section 2 we give a brief introduction to the dynamics of massive branons. 
Section 3 contains a summary of the main steps used in the standard 
calculations of relic abundances generated by the freeze-out phenomenon in an
expanding universe. In section 4, we give our results for the thermal
averages of branon annihilation cross sections into SM particles.
Section 5 is devoted to the study of the limits on the branon mass and 
on the brane tension scale from the cosmological dark matter abundances.
In section 6, after  reviewing  the limits imposed by nucleosynthesis
on the number of relativistic species, we apply them to the case of branons.
Section 7 contains the calculation of the rate of energy loss from a supernova
core in the form of branons and the corresponding limits on $f$ and $M$. 
Finally, section 8 includes the main conclusions of the paper. In an Appendix
we have included the explicit formulas for the creation and annihilation
cross sections  
of branons pairs.

\section{The branon field}
In this section we will briefly review the main properties of massive
brane fluctuations (see \cite{DoMa,BSky,ACDM} for a more detailed 
description). 
We will consider a single-brane model in large extra dimensions. 
Our four-dimensional space-time $M_4$ is embedded in a
$D$-dimensional bulk space which, for simplicity, we will assume to
be of the form $M_D=M_4\times B$. The $B$ space is a given
N-dimensional compact manifold, so that $D=4+N$. The brane lies
along $M_4$ and we neglect its contribution to the bulk
gravitational field. The coordinates parametrizing the points in
$M_D$ will be denoted by $(x^{\mu},y^m)$, where the different
indices run as $\mu=0,1,2,3$ and $m=1,2,...,N$. The bulk space
$M_D$ is endowed with a metric tensor which we will denote by
$G_{MN}$, with signature $(+,-,-...-,-)$. For simplicity, we will
consider the following ansatz:
\begin{eqnarray}
 G_{MN}&=&
\left(
\begin{array}{cccc}
\tilde g_{\mu\nu}(x,y)&0\\ 0&-\tilde g'_{mn}(y)
\end{array}\right).
\end{eqnarray}
The position of the brane in the bulk can be parametrized as
$Y^M=(x^\mu, Y^m(x))$, with $M=0,\dots, 3+N$ and
where we have chosen the bulk coordinates
so that the first four are identified with the space-time brane
coordinates $x^\mu$. We assume the brane to be created at a
certain point in $B$, i.e. $Y^m(x)=Y^m_0$ which corresponds to its
ground state. We will also assume that $B$ is a homogeneous space, so 
that brane fluctuations can be written in terms of properly normalized
coordinates in the extra space: $\pi^\alpha(x)=f^2 Y^\alpha(x)$, 
$\alpha=1,\dots, N$. The induced metric on the brane in its ground 
state is simply given by the
four-dimensional components of the bulk space metric, i.e.
$g_{\mu\nu}=\tilde g_{\mu\nu}=G_{\mu\nu}$. However, when brane
excitations  are present, the induced metric is given by
\begin{eqnarray}
g_{\mu\nu}=\partial_\mu Y^M\partial_\nu Y^N G_{MN}(x,Y(x)) =\tilde
g_{\mu\nu}(x,Y(x))-\partial_{\mu}Y^m\partial_{\nu}Y^n\tilde
g'_{mn}(Y(x))
\label{induced}
\end{eqnarray}
The contribution of branons to the
induced metric  is then obtained expanding (\ref{induced})
around the ground state \cite{DoMa,BSky,ACDM}:
\begin{equation}
g_{\mu\nu}=
\tilde g_{\mu\nu}-\frac{1}{f^4}\delta_{\alpha\beta}\partial_{\mu}\pi^\alpha
\partial_{\nu}\pi^\beta
+\frac{1}{4f^4}\tilde g_{\mu\nu}M_{\alpha\beta}^2\pi^\alpha\pi^\beta
+\dots
\end{equation}
Branons are the mass eigenstates of the brane 
fluctuations in the extra-space directions. The branon mass matrix 
$M_{\alpha\beta}$ 
is determined by the metric properties of the bulk space and, 
in the absence of a general model for the
bulk dynamics, we will consider its elements as free parameters 
(for an explicit construction see \cite{Andrianov}). Therefore,
branons are massless
only in highly symmetric cases \cite{DoMa,BSky,ACDM}.

Since in the limit in which gravity decouples $M_D\rightarrow \infty$,
branon fields still survive \cite{Contino}, 
branon effects can be studied independently of gravity. 
The mechanism responsible for the creation of the brane is
in principle unknown, and therefore we will assume that the 
brane dynamics can be described by a low-energy effective action
derived from the Nambu-Goto action
\cite{DoMa}. 
 Also, 
branon couplings to the SM fields can be obtained from the SM action 
defined on a curved background given by the
induced metric (\ref{induced}), and expanding in branon fields. 
Thus, the complete action, up to second
order in $\pi$ fields, contains the SM terms, the kinetic term for the
branons and the interaction terms between the SM particles and the
branons:
\begin{eqnarray}
S_B&=& \int_{M_4}d^4x\sqrt{g}[-f^4+ {\mathcal L}_{SM}(g_{\mu\nu})]
\nonumber\\
&=&\int_{M_4}d^4x\sqrt{\tilde g}\left[-f^4+ {\mathcal L}_{SM}
( \tilde g_{\mu\nu})  +
\frac{1}{2}\tilde g^{\mu\nu}\delta_{\alpha\beta}\partial_{\mu}
\pi^\alpha
\partial_{\nu}\pi^\beta-\frac{1}{2}M^2_{\alpha\beta}
\pi^\alpha\pi^\beta\right.
\nonumber\\
&+&
\left.\frac{1}{8f^4}(4\delta_{\alpha\beta}\partial_{\mu}\pi^\alpha
\partial_{\nu}\pi^\beta-M^2_{\alpha\beta}\pi^\alpha\pi^\beta
\tilde g_{\mu\nu})
T^{\mu\nu}_{SM}(\tilde g_{\mu\nu}) \right]
+\dots \label{lag}
\end{eqnarray}
where $T^{\mu\nu}_{SM}(\tilde g_{\mu\nu})$ is the conserved
energy-momentum tensor of the Standard Model evaluated in the
background metric.

\begin{eqnarray}
T^{\mu\nu}_{SM}=-\left(\tilde g^{\mu\nu}{\mathcal L}_{SM}
+2\frac{\delta {\mathcal L}_{SM}}{\delta \tilde
g_{\mu\nu}}\right) 
\end{eqnarray}

It is interesting to note that under a parity transformation on
the brane, the branon field changes sign if the number of spatial
dimensions of the brane is odd, whereas it remains unchanged for even
dimensions. Accordingly, branons on a 3-brane are pseudoscalar
particles. This implies that if we want to preserve parity on the
brane, terms in the effective Lagrangian with an odd number of 
branons would be forbidden.

The quadratic expression
in (\ref{lag}) is valid for any internal $B$ space,
regardless of the particular form of the metric
 $\tilde g'_{mn}$. In fact the low-energy effective lagrangian 
is model independent and is parametrized only by 
the number
of branon fields, their masses and the brane tension. 
The dependence on the
geometry of the extra dimensions will appear at higher orders.
These effective couplings thus provide the necessary tools to compute
cross sections and expected rates of events involving branons 
in terms of $f$ and the branons masses only.

From the previous expression, we see that since branons interact
by pairs with the SM particles, they are necessarily stable. In 
addition, their couplings are suppressed by the brane tension 
$f^4$, which means that they could be weakly interacting, and
finally, according to our previous discussion, in general, they are
expected to be massive. As a consequence their freeze-out temperature
can be relatively high, which implies that their relic abundances can
be cosmologically important. 

\section{Relic branon abundances}
In order to calculate the thermal relic branon abundance, we will
use the standard techniques given in \cite{Kolb,relic} in two 
limiting cases,
either  relativistic (hot) or non-relativistic (cold) 
branons at decoupling. In this section we will review the basic steps 
of the calculation method. 

The evolution of the number density $n_\alpha$ of branons
 $\pi^\alpha$,
 $\alpha=1,\dots , N$ with $N$ the number of different types
of branons, interacting with SM particles in an expanding universe
is given by the Boltzmann equation:
\begin{eqnarray}
\frac{dn_\alpha}{dt}=-3Hn_\alpha-\langle \sigma_A v\rangle
(n_\alpha^2 -(n_\alpha^{eq})^2)\label{Boltzmann}
\end{eqnarray}
where
\begin{eqnarray}
\sigma_A=\sum_X \sigma(\pi^\alpha\pi^\alpha\rightarrow X)
\end{eqnarray}
is the total
annihilation cross section of branons into SM particles $X$
summed over final states. The
$-3Hn_\alpha$
term, with $H$ the Hubble parameter,
takes into account the dilution of the number density due to
the universe expansion. These are the only terms which
could change the number density of branons. In fact,
since branons are stable they do not decay into other
particles and since they interact always by pairs the
conversions like $\pi^\alpha X \rightarrow \pi^\alpha Y$ do not
change their number. Notice that we are considering only the
lowest order Lagrangian and assuming that all the
branons have the same mass. This implies that each branon species
evolves independently. Therefore in the
following we will drop the $\alpha$
index.

The thermal average $\langle \sigma_A v\rangle$ 
 of the total annihilation cross section
times the relative velocity is given by:
\begin{eqnarray}
\langle \sigma_A v\rangle=\frac{1}{n_{eq}^2}\int
\frac{d^3p_1}{(2\pi)^3}
\frac{d^3p_2}{(2\pi)^3} f(E_1) f(E_2)\frac{w(s)}{E_1 E_2}
\end{eqnarray}
where:
\begin{eqnarray}
w(s)=E_1 E_2 \sigma_A v_{rel}=\frac{s\sigma_A}{2}
\sqrt{1-\frac{4 M^2}{s}}
\end{eqnarray}
The Mandelstam variable $s$ can be written in terms of the
components of the four momenta of the two branons $p_1$ and
$p_2$ as $s=(p_1+p_2)^2=2(M^2+E_1E_2-\vert \vec p_1\vert
\vert \vec p_2\vert \cos\theta)$.
Assuming vanishing chemical potential, the branon distribution functions are:
\begin{eqnarray}
f(E)=\frac{1}{e^{E/T} + a}
\end{eqnarray}
with $a=0$ for Maxwell-Boltzmann and $a=-1$ for Bose-Einstein.
In the case of non-relativistic relics $T\ll 3M$, the Maxwell-Boltzmann
distribution is a good approximation and we will use it for
simplicity instead of  Bose-Einstein.
Finally, the equilibrium abundance is given by:
\begin{eqnarray}
n_{eq}=\int \frac{d^3p}{(2\pi)^3} f(E)
\end{eqnarray}
From (\ref{lag}),
the thermal average will include, to leading order, annihilations 
into all 
the SM particle-antiparticle pairs.
If the universe temperature is above the
QCD phase transition ($T>T_c$),
we consider  annihilations into quark-antiquark and gluons pairs. 
If $T<T_c$  we 
include annihilations
into light hadrons. For the sake of definiteness we will take a
critical temperature $T_c\simeq 170$ MeV and a Higgs mass $m_H\simeq
125$ GeV, although the final results are not very sensitive to the
concrete value of these parameters.

In order to solve the Boltzmann equation we introduce the
new variables: $x=M/T$ and $Y=n/s$ with $s$ the universe
entropy density. We will assume that the total entropy 
of the universe is conserved, i.e. $S=a^3 s=\mbox{const}$,
where $a$ is the scale
factor of the universe and we will make use of the Friedmann equation:
\begin{eqnarray}
H^2=\frac{8\pi}{3M_P^2}\rho
\end{eqnarray}
where the energy density in a radiation dominated universe
is given by:
\begin{eqnarray}
\rho=g_{eff}(T)\frac{\pi^2}{30}T^4
\end{eqnarray}
In a similar way, the entropy density
reads:
\begin{eqnarray}
s=h_{eff}(T)\frac{2\pi^2}{45}T^3
\end{eqnarray}
where $g_{eff}(T)$ and $h_{eff}(T)$ denote the effective
number of relativistic degrees of freedom contributing
to the energy density and the entropy density respectively
at temperature $T$ ($T$ being the temperature of the photon
background). Notice that for $T>$ MeV we have
$h_{eff}\simeq g_{eff}$.
Using these expressions we get:
\begin{eqnarray}
\frac{dY}{dx}=-\left(\frac{\pi M_P^2}{45}\right)^{1/2}
\frac{h_{eff}M}{g_{eff}^{1/2}x^2}\langle\sigma_A v
\rangle(Y^2-Y_{eq}^2)
\label{YBoltz}
\end{eqnarray}
where we have ignored the possible derivative terms
$dh_{eff}/dT$.

The qualitative behaviour of the
solution of this equation goes as follows:
if the annihiliation rate defined as
$\Gamma_A=n_{eq}\langle\sigma_A v
\rangle$  is larger than the expansion rate of the universe
$H$ at a given $x$, then $Y(x)\simeq Y_{eq}(x)$, i.e., the
branon abundance follows the equilibrium
abundances. However, since  $\Gamma_A$ decreases
with the temperature, it eventually
becomes  similar to $H$  at some point $x=x_f$. From that
time on branons are decoupled from the rest of matter or
radiation
in the universe and its abundance remains frozen, i.e.
$Y(x)\simeq Y_{eq}(x_f)$ for $x\geq x_f$. 
For relativistic (hot) particles, 
the equilibrium abundance reads:
\begin{eqnarray}
Y_{eq}(x)=\frac{45\zeta(3)}{2\pi^4}\frac{1}{h_{eff}(x)},
\;\;\;\;\; (x \ll 3)\label{hot}
\end{eqnarray}
whereas for cold relics:
\begin{eqnarray}
Y_{eq}(x)=\frac{45}{2\pi^4}\left(\frac{\pi}{8}
\right)^{1/2}x^{3/2}\frac{1}{h_{eff}(x)}\,e^{-x},
\;\;\;\;\; (x \gg 3)\label{cold}
\end{eqnarray}
We see that for hot branons the equilibrium abundance is
not very sensitive to the value of $x$. In the case of cold
relics however, $Y_{eq}$ decreases exponentially
with the temperature,   which implies that
the sooner the decoupling occurs the larger the relic
abundance.

Let us first consider the simple case of  hot branons.
Since its equilibrium abundance depends
on $x_f$ only through $h_{eff}(x_f)$, the relic abundance
is not very sensitive to the exact time of decoupling.
In this case, in order to calculate the
decoupling temperature $T_f=M/x_f$,
it is a good approximation to use the condition $\Gamma_A=H$.
From the explicit expression of the Hubble parameter
in a radiation dominated universe we have:
\begin{eqnarray}
H(T_f)=1.67\, g_{eff}^{1/2}(T_f) \frac{T_f^2}{M_P}=\Gamma_A(T_f)
\label{Hubble}
\end{eqnarray}
which can be solved explicitly for $T_f$, expanding $\Gamma_A(T_f)$
for $T_f\gg M/3$.
Once we know $x_f$, the relic abundance today ($Y_{\infty}\simeq Y(x_f)$) is given by
(\ref{hot}).
From this expression we can obtain the current number density
of branons and the corresponding energy density which is given by:
\begin{eqnarray}
\Omega_{Br} h^2=7.83 \cdot 10^{-2} \frac{1}{h_{eff}(x_f)}
\frac{M}{\mbox{eV}}\label{eV}
\end{eqnarray}

The calculation of the decoupling temperature in the
case of cold branons is more involved. The well-known
result is given by:
\begin{eqnarray}
x_f=\ln\left(\frac{0.038\, c\,(c+2) M_P M \langle\sigma_A v
\rangle}{g_{eff}^{1/2}\,x_f^{1/2}}\right)
\label{xf}
\end{eqnarray}
where $c\simeq 0.5$ is obtained from the numerical solution of
the Boltzmann equation.
This equation can be solved iteratively. The corresponding 
energy fraction reads:
\begin{eqnarray}
\Omega_{Br} h^2=8.77 \cdot 10^{-11} \mbox{GeV}^{-2}
\frac{x_f}{g_{eff}^{1/2}}\left(\sum_{n=0}^\infty
\frac{c_n}{n+1}x_f^{-n}\right)^{-1}
\label{coldomega}
\end{eqnarray}
where we have expanded $\langle \sigma_A v\rangle $
in powers of $x^{-1}$ as:
\begin{eqnarray}
\langle \sigma_A v\rangle =\sum_{n=0}^\infty c_n x^{-n}
\end{eqnarray}
 Notice that in general,
$Y_\infty \propto 1/\langle \sigma_A v\rangle$, i.e.  the
weaker the cross section the larger the relic abundance.
This
is the expected result, since, as commented before
the sooner the decoupling occurs, the larger the relic
abundance, and decoupling occurs earlier as we decrease
the cross section. Therefore the cosmological bounds
work in the opposite way as compared to those coming from
colliders. Thus, a bound such as
 $\Omega_{Br} < \Od(1)$ translates into 
a lower limit for the cross sections and not into an upper limit
as those obtained from non observation in colliders.

In the following we apply the previous formalism to obtain
the relic abundance of branons $\Omega_{Br} h^2$, both when
they are relativistic and non-relativistic at decoupling.
For that purpose we need to evaluate the thermal averages
$\langle \sigma_A v\rangle $ for the annihilation of branons
into photons, massive $W^{\pm}$ and $Z$ gauge bosons,
three massless neutrinos, charged leptons, quarks and gluons (or 
light hadrons) and a real scalar
Higgs field, in terms of the brane tension
$f$ and the branon mass $M$.

\section{Branon annihilation cross sections: thermal average}

We give the results for the
different channels contributing to
the thermal average of the annihilation
cross section $\langle \sigma_A v\rangle$ of branons
into SM particles. The explicit production and annihiliation cross
section can be found in the Appendix. 
For cold relics, we have expanded the 
expressions for each particle species in powers
of $1/x$ as follows:
\begin{eqnarray}
\langle \sigma_A v\rangle=c_0
+c_1\frac{1}{x}+c_2\frac{1}{x^2}+
\Od(x^{-3})
\end{eqnarray}

In the hot branons case, we  give the results for the different
contributions to the decay rate $\Gamma_A=n_{eq}\langle\sigma_A v
\rangle$, where we have considered the ultrarrelativistic
limit for the branons, i.e. $M=0$. For massive SM particles, the final 
expressions cannot be given in closed form. Therefore, in this
section, in order to show the high-temperature behaviour, 
we only give the results for  fermions, gauge bosons and
scalars in the limit in which their masses vanish. 
Also in this case, we have used the Bose-Einstein form as the equilibrium
distribution.

\subsection{Dirac fermions}
\subsubsection*{$x\gg 3$ (Cold)}
%\begin{eqnarray}
%\langle \sigma_A^{Dirac} v\rangle=c_0
%+c_1\frac{1}{x}+c_2\frac{1}{x^2}+
%\Od(x^{-3})
%\end{eqnarray}
%where
\begin{eqnarray}
c_0&=&\frac{1}{16\pi^2f^8}M^2 m_\psi^2(M^2-m_\psi^2)
\sqrt{1-\frac{m_\psi^2}{M^2}}\\
c_1&=&\frac{1}{192\pi^2f^8}M^2 m_\psi^2(67M^2-31m_\psi^2)
\sqrt{1-\frac{m_\psi^2}{M^2}}\\
c_2&=&\frac{1}{7680\pi^2f^8}\frac{M^2}{M^2-m_\psi^2}
(17408M^6+13331M^4 m_\psi^2\nonumber \\
&-&46606M^2 m_\psi^4
+18927 m_\psi^6)
\sqrt{1-\frac{m_\psi^2}{M^2}}
\end{eqnarray}
Notice that this expansion is not valid near SM particles
thresholds, i.e. for branon masses close to some SM particle mass.
In addition, since the $c_0$ coefficient is different from
zero, annihilation will mainly take place through $s$-wave. 
\subsubsection*{$x\ll 3$ (Hot)}
For massless fermions, we obtain:
\begin{eqnarray}
\Gamma_A^{Dirac}=\frac{8\pi^9 T^9}{297675 \zeta(3)f^8}+\Od(x)
\end{eqnarray}
\subsection{Massive gauge field}
\subsubsection*{$x\gg 3$ (Cold)}
%\begin{eqnarray}
%\langle \sigma_A^{Z} v\rangle=c_0
%+c_1\frac{1}{x}+c_2\frac{1}{x^2}+
%\Od(x^{-3})
%\end{eqnarray}
%where
\begin{eqnarray}
c_0&=&
\frac{M^2\,{\sqrt{1 - \frac{{m_Z}^2}{M^2}}}\,
    \left( 4\,M^4 - 4\,M^2\,{m_Z}^2 + 3\,
{m_Z}^4 \right) }{64\,f^8\,{\pi }^2}\\
c_1&=&\frac{M^2\,{\sqrt{1 - \frac{{m_Z}^2}{M^2}}}\,
    \left( 364\,M^6 - 584\,M^4\,{m_Z}^2 + 349\,M^2\,
{m_Z}^4 - 93\,{m_Z}^6 \right) }{768\,
    f^8\,\left( M^2 - {m_Z}^2 \right) \,{\pi }^2}\nonumber\\
c_2&=&\frac{M^2\,{\sqrt{1 - \frac{{m_Z}^2}{M^2}}}\,
     }{30720\,f^8\,
    {\left( M^2 - {m_Z}^2 \right) }^2\,{\pi }^2}
\left( 415756\,M^8 - 755844\,M^6\,{m_Z}^2\right. \nonumber\\
&+& \left. 356541\,M^4\,
{m_Z}^4 -
      76294\,M^2\,{m_Z}^6 + 56781\,{m_Z}^8 \right)\nonumber
\end{eqnarray}
Again this expansion is not valid near SM particles
thresholds, and the leading term corresponds to the $s$-wave.
\subsubsection*{$x\ll 3$ (Hot)}
In the limit where $T\gg m_Z$, one can obtain the following expression:
\begin{eqnarray}
\Gamma_A^{Z}=\frac{8\pi^9
T^9}{99225 \zeta(3)f^8}+\Od(x)
\end{eqnarray}
\subsection{Massless gauge field}
\subsubsection*{$x\gg 3$ (Cold)}
%\begin{eqnarray}
%\langle \sigma_A^{\gamma} v\rangle=c_0
%+c_1\frac{1}{x}+c_2\frac{1}{x^2}+ \Od(x^{-3})
%\end{eqnarray}
%where
\begin{eqnarray}
c_0&=&0\\
c_1&=&0\nonumber\\
c_2&=&\frac{68\,M^6}{15\,f^8\,{\pi }^2}\nonumber
\end{eqnarray}
In this case, $c_0=c_1=0$ and the leading term corresponds to
$d$-wave annihilation.
\subsubsection*{$x\ll 3$ (Hot)}
\begin{eqnarray}
\Gamma_A^{\gamma}=\frac{16\pi^9 T^9}{297675 \zeta(3)f^8}+\Od(x)
\end{eqnarray}
\subsection{Complex scalar field}
\subsubsection*{$x\gg 3$ (Cold)}
%\begin{eqnarray}
%\langle \sigma_A^{\Phi} v\rangle=c_0
%+c_1\frac{1}{x}+c_2\frac{1}{x^2}+
%\Od(x^{-3})
%\end{eqnarray}
%where
\begin{eqnarray}
c_0&=&\frac{M^2\,{\left( 2\,M^2 + {m_\Phi}^2 \right) }^2\,
{\sqrt{1 - \frac{{m_\Phi}^2}{M^2}}}}{32\,f^8\,{\pi }^2}\\
c_1&=&\frac{M^2\,\left( 2\,M^2 + {m_\Phi}^2 \right)
\,{\sqrt{1 - \frac{{m_\Phi}^2}{M^2}}}\,
    \left( 182\,M^4 - 115\,M^2\,{m_\Phi}^2
- 31\,{m_\Phi}^4 \right) }{384\,f^8\,
    \left( M^2 - {m_\Phi}^2 \right) \,{\pi }^2}\nonumber\\
c_2&=&\frac{M^2\,{\sqrt{1 - \frac{{m_\Phi}^2}{M^2}}}\,
     }{5120\,f^8
\,{\left( M^2 - {m_\Phi}^2 \right) }^2\,{\pi }^2}
\left( 92164\,M^8 - 123556\,M^6\,{m_\Phi}^2 \right.\nonumber \\
&+& \left. 12269\,M^4\,{m_\Phi}^4 + 9754\,M^2\,{m_\Phi}^6 +
      6309\,{m_\Phi}^8 \right)\nonumber
\end{eqnarray}
We find the same problems near SM particles
thresholds. The dominant contribution is the $s$-wave.
\subsubsection*{$x\ll 3$ (Hot)}
For massless scalar, one can obtain the following expression:
\begin{eqnarray}
\Gamma_A^{\Phi}=\frac{16\pi^9 T^9}{297675 \zeta(3)f^8}+\Od(x)
\end{eqnarray}
For real scalar fields, the above results should be divided by two.

Notice  that for conformal matter the leading contribution is the $d$-wave.
This explains why 
in the massless limit for fermions, the leading contribution 
is no longer the $s$-wave, but the $d$-wave, whereas for massless
scalars  or taking the $m_Z\rightarrow 0$ limit
for massive gauge bosons, the $s$- and $p$-waves
survive.

Concerning the validity of the above results, 
in order to avoid the mentioned problems of the
Taylor expansion near SM thresholds, we have taken branon masses
sufficiently separated from SM particles masses where the usual
treatment is adequate \cite{Kolb,relic}. Such treatment 
is known to introduce errors of the order of 10$\%$ in the
relic abundances. In addition, 
coannihilation effects are absent in this case  since there are no
slightly heavier particles which eventually could decay into the
lightest branon.

\section{Cosmological bounds from the dark matter energy density}
For cold branons, once we know the $c_n$ coefficients for the total
cross section, we can compute the freeze-out
value $x_f$ from (\ref{xf}) and the relic contribution
to the energy density of the universe $\Omega_{Br}h^2$
from (\ref{coldomega}) in terms of $f$ and $M$. Imposing the
observational limit on the total dark matter energy density
from WMAP: $\Omega_{Br}h^2 < 0.129 - 0.095$ 
at the 95$\%$ C.L. which
corresponds to $\Omega_M=0.23 \pm 0.08$ and $h=0.79-0.65$ 
\cite{WMAP}, 
we obtain the exclusion plots in Figs. 1 and 6.
\begin{figure}[h]
{\epsfxsize=12.0 cm \epsfbox{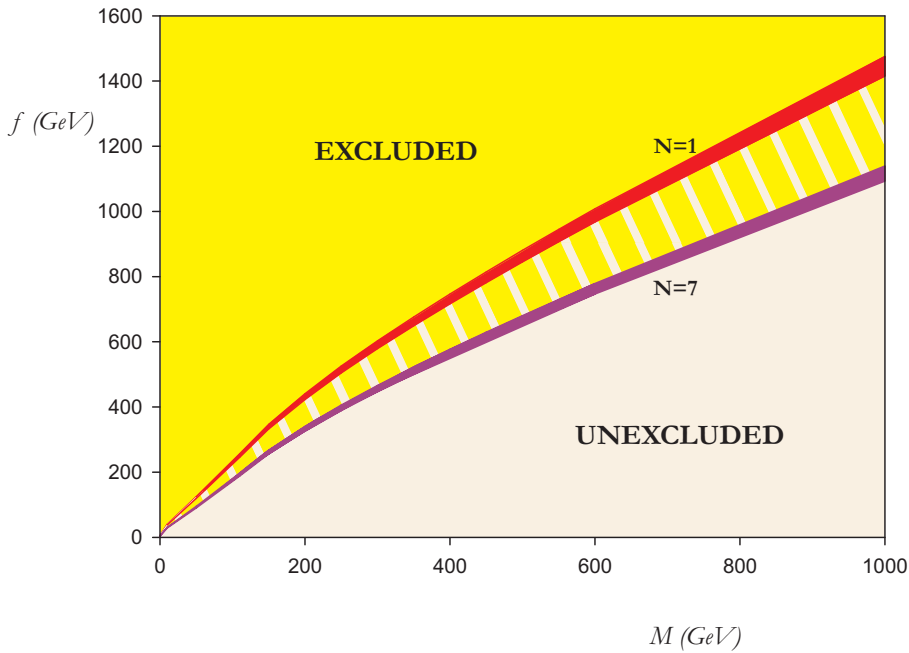}}
\caption{Exclusion plot from cold relics abundance. The thick lines 
correspond to 
$\Omega_{Br}h^2=0.129-0.095$ for $N=1$ and $N=7$. Therefore the areas above
the curves corresponding
to $\Omega_{Br}h^2>0.129$ are excluded. The striped region contains the 
curves corresponding to $1<N<7$.}
\end{figure}

Notice that in Fig.6 we have plotted the $x_f=3$ curve,
which limits the range of validity of the cold relic
approximation.
Therefore, the excluded region is that
between the two curves.
It is also important to note that for those values
of the parameters on the solid line, branons
would constitute all the dark matter in the universe.

For hot branons, 
we have computed numerically the total annihilation
rate  into SM particles $\Gamma_A$. Using 
equation (\ref{Hubble}), we can find the freeze-out
temperature $T_f$ in terms of the brane tension scale $f$. 
Approximately, 
the relation between the logarithms of these quantities is linear:
$\log_{10} (f/1 \mbox{GeV}) 
\simeq(7/8)\log_{10}(T_f/1 \mbox{ GeV}) + 2.8$. This expression 
is almost independent of the number of branons.
From the numerical values of $T_f$ in terms of $f$, it is possible to
obtain $\Omega_{Br}h^2$ from (\ref{eV}). In this case we have considered
two kinds of limits. On one hand those coming
from the total dark matter of the universe $\Omega_{Br}h^2 < 0.129 - 0.095$
in Fig. 2. On the other hand, more constraining limits on the hot dark matter
energy density  
can be derived from a combined analysis of the data from 
WMAP, CBI, ACBAR, 2dF and Lyman $\alpha$ \cite{WMAP}. The bound reads
$\Omega_{Br}h^2<0.0076$ at the 95$\%$C.L. and it is obtained thanks to the
fact that hot dark matter is able
to cluster on large scales but free-streaming reduces the power on small 
scales, changing the shape of the matter power spectrum. In Fig. 3 we have 
plotted the corresponding limits in the $f-M$ plane. Notice the abrupt
jump around $f\simeq 60$ GeV in Figs. 2 and 3 (and also in Fig. 4).
This $f$ value corresponds to a decoupling temperature of $T\simeq 170$ MeV 
which is the assumed value for the QCD phase transition. Thus, this jump
is due to the sudden growth in the number of  effective degrees of freedom
when passing from the hadronic  to the quark-gluon plasma phase.
The exclusion areas depend on the number of branon species 
and we have plotted them
for $N=1,2,3,7$ in Figs. 2,3.
\begin{figure}[h] {\epsfxsize=12.0 cm \epsfbox{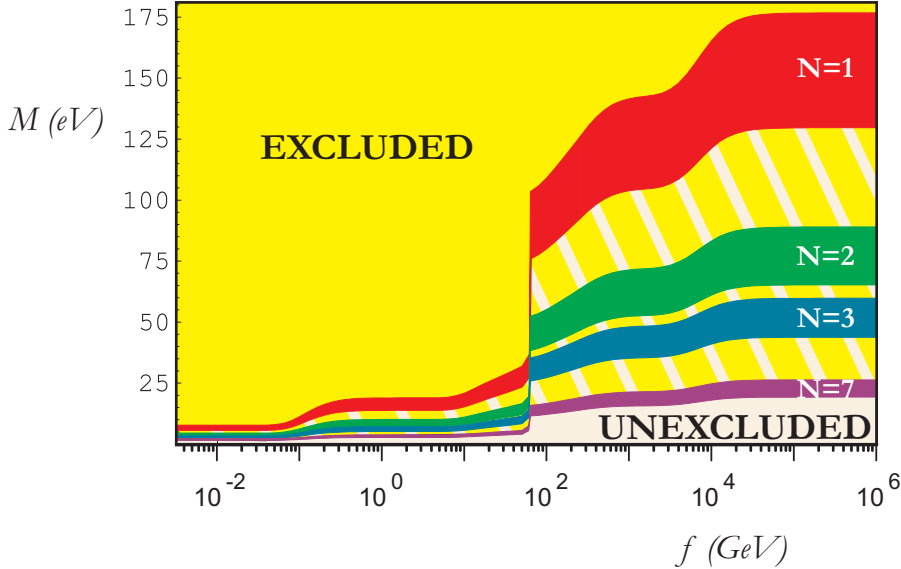}}
\caption{Exclusion plot for hot relics in models
with $N=1,2,3,7$ branons. For a given $N$, the shaded area corresponds
to the total dark matter limit 
$\Omega_{Br}h^2 = 0.129 - 0.095$, therefore the region above such
area is excluded by branon overproduction.}
\end{figure}
The validity of the previous limits requires that  branons were
relativistic particles at  freeze-out. Therefore we require
$x_f \ll 3$, which implies that the bounds do not work for
$f<10^{-4}$ GeV. The curve $x_f=3$ in the hot relic case is also 
plotted in  Fig. 6.

As commented in the introduction, in all the previous calculations we are
assuming, apart from $f\ll M_D$, that  the evolution of
the universe is standard up to a temperature around $f$.
In fact, the 
effective Lagrangian (\ref{lag}) is only valid at low energies 
relative
to  $f$ and therefore, it is this scale what fixes the range of
validity of the results. We have checked that our calculations are consistent 
with these
assumptions 
since the decoupling temperatures   
are always smaller than $f$ in the allowed regions in Fig. 6.
\begin{figure}[h]{\epsfxsize=12.0 cm \epsfbox{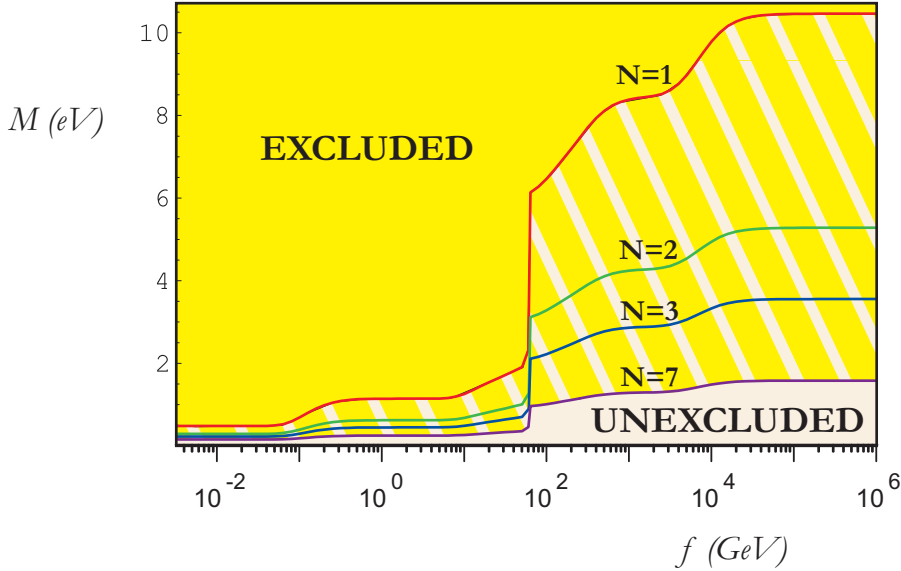}}
\caption{Exclusion plot for hot relics in models
with $N=1,2,3,7$ branons. For a given $N$, the curve corresponds
to the hot dark matter limit 
$\Omega_{Br}h^2=0.0076$, therefore the region above such
curve is excluded by hot branon overproduction.}
\end{figure}

\section{Bounds from nucleosynthesis}
As commented above, one of the most successful predictions of the 
standard cosmological
model is the relative abundance of the light elements.
The calculated abundances are very sensitive
to certain cosmological parameters, in particular it has been shown
that the production of  $^4$He increases with increasing rate
of expansion $H$. From (\ref{Hubble}) we see that the Hubble parameter
depends on the effective number of relativistic degrees of
freedom $g_{eff}$. Usually, this number is parametrized in terms
of the effective number of neutrino species $N_\nu=3+\Delta N_\nu$ as:
\begin{eqnarray}
g_{eff}(T\sim \mbox{MeV})= g_{eff}^{SM}+g_{eff}^{new}\leq 
10.75+\frac{7}{4}\Delta N_\nu
\label{neutrino}
\end{eqnarray}
where $T\sim \mbox{MeV}$ corresponds to the universe temperature
during nucleosynthesis. In the SM with three massless neutrino
families, we have $g_{eff}^{SM}(T\sim \mbox{MeV})=10.75$
corresponding to the photon field, the three neutrinos and the
electron field. In (\ref{neutrino}), 
in order to avoid deviations of the 
predicted abundances from observations, the conservative
limit $\Delta N_\nu = 1$ for the contribution from
new physics is usually imposed \cite{Kolb}, i.e. 
there could be only one new type of light neutrino. 

Including branons, the number of relativistic degrees of freedom at
a given temperature $T$ is given by:
\begin{eqnarray}
g_{eff}(T)=g_{eff}^{SM}(T)+N\left(\frac{T_B}{T}\right)^4
\label{geff}
\end{eqnarray}
where $g_{eff}^{SM}(T)$ is the contribution from the SM particles,
$T_B$ denotes the  temperature of the cosmic branon  brackground
and we are assuming that there are no additional new particles. 
If branons are not decoupled
at a temperature $T$ then $T_B=T$, i.e. they have the
same temperature as the photons. On the other hand,
if they
are already decoupled then its temperature will be in general lower
 than that of the photons.
In order to calculate it,
we use
the fact that the universe expansion is adiabatic. Let us write:
\begin{eqnarray}
h_{eff}(T)=h_{eff}^{SM}(T)+N\left(\frac{T_B}{T}\right)^3
\end{eqnarray}
where $h_{eff}^{SM}(T)$ includes only the contributions from SM particles.
If at some time between branon freeze-out and nucleosynthesis, 
some other particle species becomes non-relativistic while still
in thermal equilibrium with the photon background, then
its entropy is transferred to the photons, but not
to the branons which are already decoupled. Thus, the entropy
transfer increases the photon temperature relative
to the branon temperature.
The total entropy of particles in equilibrium with
the photons remains constant i.e.:
\begin{eqnarray}
h_{eff}^{eq} a^3 T^3=\mbox{constant}
\end{eqnarray}
and since the number of relativistic
degrees of freedom $h_{eff}^{eq}$ has decreased, then
$T$ should increase
with respect to $T_B$.
Thus, we find:
\begin{eqnarray}
\frac{g_{eff}^{eq}(T_{f,\,B})}{g_{eff}^{eq}(T)}=\frac{T^3}{T_B^3}
\end{eqnarray}
where $T_{f,\,B}$ is the branon freeze-out temperature 
 and we have used the fact that for particles
in equilibrium with the photons
$g_{eff}^{eq}=h_{eff}^{eq}$. Using (\ref{geff}), 
 we can
set the following limit on the number of
massless
branon species $N$:
\begin{eqnarray}
\frac{7}{4}\Delta N_\nu\geq N\left(\frac{T_B}{T_{nuc}}\right)^4=
N\left(\frac{g_{eff}^{eq}(T_{nuc})}
{g_{eff}^{eq}(T_{f,\,B})}\right)^{4/3}
\end{eqnarray}
 If branons decouple
after nucleosynthesis we get the direct limit:
\begin{eqnarray}
N\leq \frac{7}{4}\Delta N_\nu \label{N1}
\end{eqnarray}
If they decouple before, we have $g_{eff}^{eq}(T_{nuc})=10.75$,
as seen before. Accordingly, we can rewrite the
bound as:
\begin{eqnarray}
N\leq \frac{7}{4}\Delta N_\nu\left(\frac{g_{eff}(T_{f,\,B})}{10.75}\right)^{4/3}
\label{N2}
\end{eqnarray}
 Taking $\Delta N_\nu=1$, the relation between the freeze-out 
temperature $T_{f,\, B}$ and
the brane tension scale $f$ that we have obtained for 
hot relics can be used
to get limits  on the
number of branons $N$ (see Fig. 4). Thus, for $f< 10$ GeV, we
get $N\leq 1$. This result is obtained using 
eq.(\ref{N1}) for $f<3$ GeV (which corresponds to $T_{f,B}\lsim 1$ MeV)  
or eq.(\ref{N2}) otherwise. 
However the limits are  less restrictive in the range $f\simeq 10-60$ 
GeV. 
In this case we get $N \leq 3$. Above the QCD phase transition which,  
as commented before, 
corresponds
to $f\simeq 60$ GeV,  the bound 
rises so much that the
restrictions are very weak. Notice that we are taking 
$g_{eff}^{SM}(T\gsim 300 \;\mbox{GeV})=106.75$, i.e. we only include
the minimal Standard Model matter content with a Higgs doublet and three massless
neutrinos. 
\begin{figure}[h] {\epsfxsize=12.0 cm \epsfbox{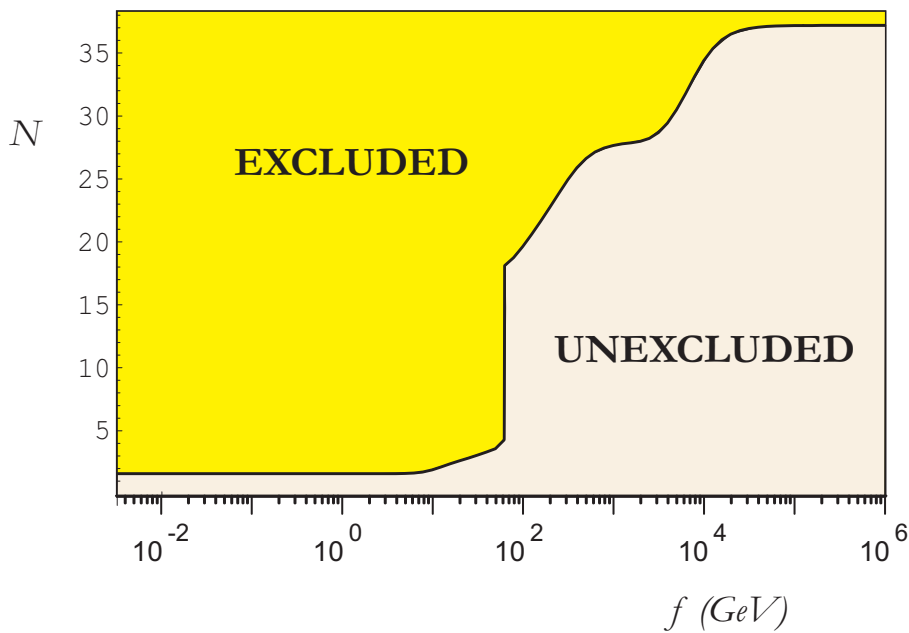}}
\caption{Restrictions from nucleosynthesis on the number of massless 
branon species
$N$ as a function of  $f$ for $\Delta N_\nu=1$.
}
\end{figure}

Concerning the value of $\Delta N_\nu$,  more constraining analysis
using only BBN suggest $\Delta N_\nu= 0.5$ \cite{Abazajian}. However, using 
also WMAP results, the constraints are $N_\nu=2.6^{+0.4}_{-0.3}$ (95$\%$ C.L.), which 
are only marginally consistent with
the LEP measurements of the number of neutrino families 
\cite{Steigman}. Such values
would severely constrain the number of any new light particles present
during nucleosynthesis. However, it has been suggested that these 
results  could be potentially affected by systematics errors in the BBN 
predictions of the primordial abundances \cite{Steigman,Olive}.

\section{Bounds from supernova SN1987A}
%\subsection{Introduction}

Important astrophysical bounds on the brane
tension scale can be obtained from the
energy loss in supernovae \cite{Kugo}. Such energy loss is carried
away essentially by light particles, i.e. photons and
neutrinos  if we restrict ourselves to 
SM particles only.
However, if the branon mass is low enough, we expect 
branons to carry some fraction of the energy, whose importance will depend on
their couplings to the SM particles. In this section, we study the
constraints on $f$ and  $M$ imposed by the cooling process of 
the neutron star in supernovae
explosions. We will perform an analysis of the energy emission rate
from the supernova core, similar to that done by Kugo and Yoshioka
for massless branons \cite{Kugo}. The aim is to extend their study to 
arbitrary $M$ and compare with the colliders and cosmological 
bounds. In their work, they consider the channel corresponding to
electron-positron pair annihilation. Although this contribution could be 
subdominant, it will allow us to get an order of magnitude estimation of the 
branon effect.

\begin{figure}[h] {\epsfxsize=12.0 cm \epsfbox{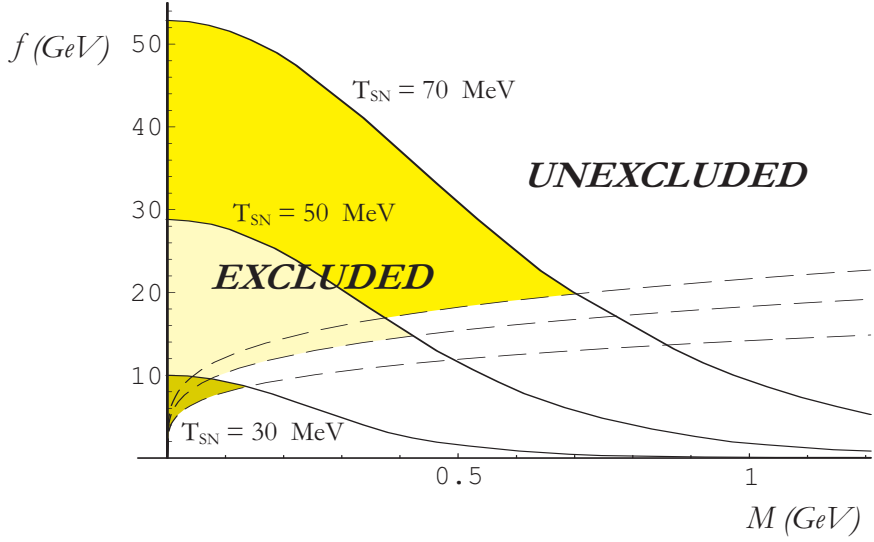}}
\caption{Exclusion regions from supernovae cooling by branons for
$T_{SN}=10,\, 50,\, 70$ MeV and $N=1$. The solid lines
come from the limits on the volume emissivity, and the dashed lines
are the $L=10$ Km limits on the branon mean free path.}
\label{SN}
\end{figure}

If branons are produced in the core, they can be scattered or
absorbed again depending on their couplings to the SM particles. 
Only if the branon mean free path $L$ inside 
the neutron star is larger than the star size ($R\sim \Od (10)$ Km), they
could escape and carry the energy away. For a massive branon 
($M\gg T_{SN}$), we have $L\sim (8\pi f^8)/(M^2 T_{SN}^4 n_e)$,
where $n_e$ is the electron number density in the star. Therefore
the restrictions we will obtain will be valid typically only for 
$f\gsim 5$ GeV. These restrictions appear due to the fact that the 
emitted energy in the form of branons 
could spoil the agreement between the predictions for the
neutrino fluxes from supernova 1987A and the observations in the Kamiokande II
\cite{kamiokandeII} and IMB \cite{IMB} detectors. Branons could shorten
the duration of the neutrino signal if the energy loss rate per 
unit time and volume is $Q \gsim 5\times 10^{-30}$
GeV$^5$. In particular, the contribution of the mentioned channel to the
volume emissivity has the form:

\begin{eqnarray}
Q_{Br}(f,M)&\equiv& \int\prod_{i=1}^2 \left\{\frac{d^3k_i}{(2\pi)^3
2E_i}2f_i \right\} (E_1+E_2)2s\,\,\sigma_{e^+e^-\rightarrow
\pi\pi}(s,f,M)
\end{eqnarray}
where $i$ refers to the electron (1) and positron (2) particles,
whose masses can be neglected in the supernova core.  The
chemical potential in the Fermi-Dirac distribution function
$f_i=1/(e^{(E_i/T-\mu/T)}+1)$ can be estimated as $\mu\sim (3
\pi^2 n_e)^{1/3}$ with the number density of electrons: $n_e\sim
1.4\times 10^{-3}\; \mbox{GeV}^3$. With these assumptions, $Q_{Br}$ is given by:

\begin{eqnarray}
Q_{Br}&=& \int_0^\infty dE_1\int_{M^2/E_1}^\infty dE_2\int_{-1}^{1
- 2 M^2/(E_1E_2)} d(cos)(E_1+E_2)
\nonumber\\
&&\frac{N[2E_1E_2\left(2E_1E_2(1-\cos)-4M^2\right)]^{5/2}\left(1-\cos\right)^{3/2}}
{{(2\pi)}^5\,7680\,f^8\,\left( 1 + e^{\frac{E_1-\mu}{T}} \right)
    \,\left( 1 + e^{\frac{E_2+\mu}{T}} \right)}
\label{Q}
\end{eqnarray}

In fact, it is possible to calculate analytically the angular
integral, whereas the integral over the two energies has been
performed numerically. The corresponding constraints depend on the 
supernova temperature ($T_{SN}$)
and the number of branons ($N$). In Fig. \ref{SN} we show the limits 
on $f$ and $M$ for  $T_{SN}=30,\,50,\,70$ MeV and $N=1$.

It is interesting to note that for a branon
mass  of the order of the GeV, the restrictions on the brane
tension scale disappear even for $T_{SN}=70$ MeV due to the limitations 
in the mean free path discussed above.

\section{Conclusions}
In this work we have studied the limits that cosmology and astrophysics
impose on the brane-world scenario through the effects of massive
branons. Using the effective low-energy Lagrangian for massive branons 
interacting with the SM fields, we have computed the annihilation cross 
sections of branon pairs into SM particles. From the solutions of the 
Boltzmann equation in an expanding universe, we have studied the 
freeze-out mechanism for branons
and obtained their thermal relic abundances both for the cold and hot cases.
Comparing the results with the recent observational limits on the total and
hot dark matter energy densities, we have obtained exclusion plots in the
$f-M$ plane. Such plots are compared with the limits coming from collider
experiments and show that there are essentially two allowed regions
in Fig. 6: one with
low branon masses and large brane tensions (weak couplings) corresponding
to hot branons, and a second region with large masses and low tensions (strong
couplings) in which branons behave as cold relics. In addition, there is
an intermediate region where $f$ is comparable to $M$, which is  precisely
the region studied in \cite{CDM}, and where branons
could account for the measured cosmological dark matter.

\begin{figure}[h]
{\epsfxsize=12.0 cm \epsfbox{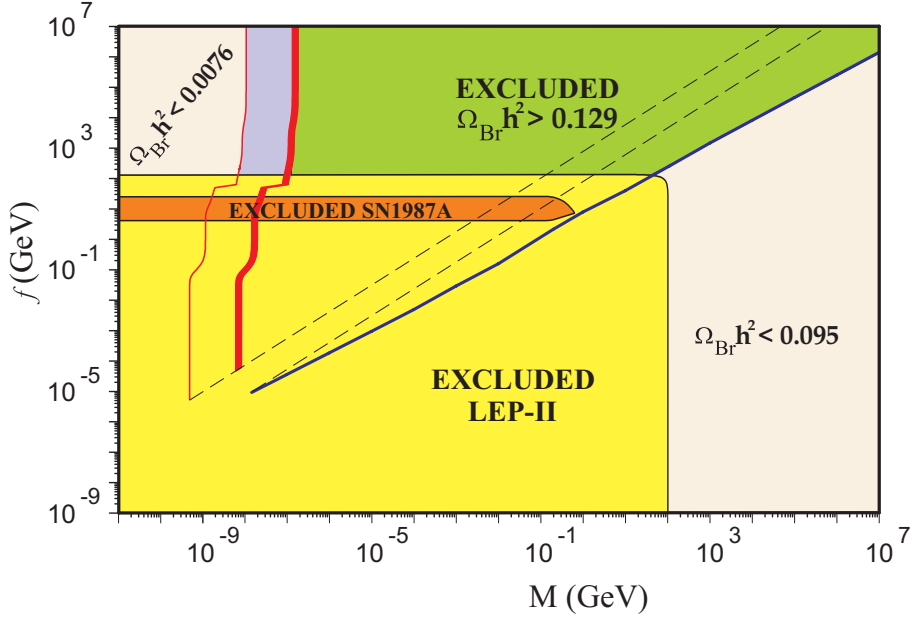}} \caption{Combined 
exclusion regions in a model with a single branon from total and hot dark
matter, LEP-II single photon events \cite{ACDM} and supernovae cooling.
The (blue) solid line on the right correspond to the cold dark matter
limit. The two (red) solid lines on the left correspond to hot dark matter: the
thicker one comes from the total dark matter range 
$\Omega_{Br}h^2=0.129-0.095$, whereas the thin one is the hot dark matter
limit $\Omega_{Br}h^2=0.0076$. The two dashed lines correspond to $x_f=3$
for hot (upper line) and cold (lower line) dark matter.
}
\end{figure}
Using the nucleosynthesis limits on the number of relativistic species, we set
a bound on the number of light branons in terms of the brane tension. We see
that if branons decouple after the QCD phase transition corresponding to
$f< 60$ GeV, the limits can be rather stringent $N\leq 3$, whereas they become
very weak otherwise.

Finally, we have analysed the possibility that massive branons could contribute
to the cooling of a supernova core. After estimating the energy loss rate, we
again get some limits on the $f$ and $M$ parameters which are compared to the
previous ones. It is shown that they are not competitive with those coming
from LEP-II.

In conclusion, cosmology imposes limits on the BWS which are complementary
to those coming from collider experiments and astrophysics. Although 
the combination of both bounds excludes an important region of the parameters
space, still there are brane-world model which could be compatible
with observations. Future hadronic colliders such as LHC, Tevatron-II or
the planned linear electron-positron colliders,  
and the possibility of detecting
dark matter branons directly \cite{CDM} or indirectly  
will allow us to explore a wider region of the parameter space. Work is
in progress in these directions.

\newpage
\appendix
\section{Branon production and 
annihilation cross sections with SM particles}

In this section we present the  branon  production and
annihilation cross sections in processes involving SM particles. The
results are presented with all the internal degrees of freedom
summed for the final particles and averaged for the initial ones. We
have used the Feynman rules given in \cite{ACDM}, where $N$ is the
number of branons.

\subsection{Scalars}

\centerline{$\sigma_1$:
$\Phi^\dagger(p_1),\Phi(p_2)\longrightarrow\pi(p_3),\pi(p_4)$}
\begin{eqnarray}
\sigma_1&=&\frac{N}{7680
f^8\pi s}\sqrt{\frac{(s-4M^2)}{(s-4m_\Phi^2)}}[-s(8m_\Phi^2+s)(s-4M^2)^2\nonumber\\
&&+(2m_\Phi^2+s)^2(23M^4-14M^2s+3s^2)].
\end{eqnarray}

\vspace{.5cm}

\centerline{$\sigma_2$: $\pi(p_1),\pi(p_2)\longrightarrow
\Phi^\dagger(p_3),\Phi(p_4)$}
\begin{eqnarray}
\sigma_2&=&\frac{1}{3840 N f^8\pi s}\sqrt{\frac{(s-4m_\Phi^2)}
{(s-4M^2)}}[-s(8m_\Phi^2+s)(s-4M^2)^2\nonumber\\
&&+(2m_\Phi^2+s)^2(23M^4-14M^2s+3s^2)].
\end{eqnarray}

The scalar results are given for a complex scalar
 such as charged pions or kaons.
For a real scalar like the Higgs field or the neutral
pion, the only change comes from the branon annihilation cross
section that should be divided by two.

\subsection{Fermions}

\centerline{$\sigma_3$:
$\psi^+(p_1),\psi^-(p_2)\longrightarrow\pi(p_3),\pi(p_4)$}
\begin{eqnarray}
\sigma_3&=&\frac{N}{30720
f^8\pi}\sqrt{(s-4M^2)(s-4m_{\psi}^2)}[(s-4M^2)^2\nonumber\\
&&+\frac{2m_{\psi}^2}{s}(23M^4-14M^2s+3s^2)].
\end{eqnarray}

\vspace{.5cm}

\centerline{$\sigma_4$: $\pi(p_1),\pi(p_2)\longrightarrow
\psi^+(p_3),\psi^-(p_4)$}

\begin{eqnarray}
\sigma_4&=&\frac{1}{3840
N f^8\pi}\sqrt{\frac{(s-4m_{\psi}^2)^3}{(s-4M^2)}}[(s-4M^2)^2\nonumber \\
&+&\frac{2m_{\psi}^2}{s}(23M^4-14M^2s+3s^2)].
\end{eqnarray}

These results are valid
for a Dirac fermion of mass $m_{\psi}$. For massless Weyl fermions, 
we should  multiply 
the branon production cross section  by two,
whereas the annihilation cross section should be divided by
the same factor of two.

\subsection{Photons}

\centerline{$\sigma_5$:
$\gamma(p_1),\gamma(p_2)\longrightarrow\pi(p_3),\pi(p_4)$}
\begin{eqnarray}
\sigma_5&=&\frac{N}{7680
f^8\pi}\sqrt{1-\frac{4M^2}{s}}s(s-4M^2)^2.
\end{eqnarray}

\vspace{.5cm}

\centerline{$\sigma_6$:
$\pi(p_1),\pi(p_2)\longrightarrow\gamma(p_3),\gamma(p_4)$}
\begin{eqnarray}
\sigma_6&=&\frac{1}{1920 N
f^8\pi}\frac{s(s-4M^2)^2}{\sqrt{1-\frac{4M^2}{s}}}.
\end{eqnarray}

\subsection{$Z$}
\centerline{$\sigma_7$:
$Z(p_1),Z(p_2)\longrightarrow\pi(p_3),\pi(p_4)$}
\begin{eqnarray}
\sigma_7&=&\frac{N}{69120 f^8\pi s} \sqrt{\frac{s-4M^2}{s-4M_Z^2}}
[3s(8M_Z^2+s)(s-4M^2)^2\nonumber\\
&&+(12M_Z^4+4sM_Z^2+s^2)(23M^4-14M^2s+3s^2)].
\end{eqnarray}

\vspace{.5cm}

\centerline{$\sigma_8$: $\pi(p_1),\pi(p_2)\longrightarrow
Z(p_3),Z(p_4)$}
\begin{eqnarray}
\sigma_8&=&\frac{1}{7680 N f^8\pi s}
\sqrt{\frac{s-4M_Z^2}{s-4M^2}}
[3s(8M_Z^2+s)(s-4M^2)^2\nonumber\\
&&+(12M_Z^4+4sM_Z^2+s^2)(23M^4-14M^2s+3s^2)].
\end{eqnarray}

\subsection{$W^\pm$}
\centerline{$\sigma_9$:
$W^\pm(p_1),W^\mp(p_2)\longrightarrow\pi(p_3),\pi(p_4)$}
\begin{eqnarray}
\sigma_9&=&\frac{N}{69120 f^8\pi s} \sqrt{\frac{s-4M^2}{s-4M_W^2}}
[3s(8M_W^2+s)(s-4M^2)^2\nonumber\\
&&+(12M_W^4+4sM_W^2+s^2)(23M^4-14M^2s+3s^2)].
\end{eqnarray}

\vspace{.5cm}

\centerline{$\sigma_{10}$: $\pi(p_1),\pi(p_2)\longrightarrow
W^\pm(p_3),W^\mp(p_4)$}
\begin{eqnarray}
\sigma_{10}&=&\frac{2}{7680 N f^8\pi s}
\sqrt{\frac{s-4M_W^2}{s-4M^2}}
[3s(8M_W^2+s)(s-4M^2)^2\nonumber\\
&&+(12M_W^4+4sM_W^2+s^2)(23M^4-14M^2s+3s^2)].
\end{eqnarray}

\subsection{Gluons}
\centerline{$\sigma_{11}$:
$g(p_1),g(p_2)\longrightarrow\pi(p_3),\pi(p_4)$}
\begin{eqnarray}
\sigma_{11}&=&\frac{N}{61440
f^8\pi}\sqrt{1-\frac{4M^2}{s}}s(s-4M^2)^2.
\end{eqnarray}

\vspace{.5cm}

\centerline{$\sigma_{12}$: $\pi(p_1),\pi(p_2)\longrightarrow
g(p_3),g(p_4)$}
\begin{eqnarray}
\sigma_{12}&=&\frac{1}{240 N
f^8\pi}\frac{s(s-4M^2)^2}{\sqrt{1-\frac{4M^2}{s}}}.
\end{eqnarray}

\end{document}